\documentclass[preprint,showpacs,preprintnumbers,amsmath,amssymb]{revtex4}

\usepackage{graphicx}
\usepackage{dcolumn}
\usepackage{bm}

\def\z{{\mathbf{z}}}

\begin{document}

\preprint{Imperial/TP/03-4/7}

\title{Disentanglement and Decoherence by Open System Dynamics}

\author{P.J.Dodd}
\author{J.J.Halliwell}%
\affiliation{Blackett Laboratory \\ Imperial College \\ London SW7
2BZ \\ UK }



\begin{abstract}
The destruction of quantum interference, decoherence, and the
destruction of entanglement both appear to occur under the same
circumstances. To address the connection between these two
phenomena, we consider the evolution of arbitrary initial states
of a two-particle system under open system dynamics described by a
class of master equations which produce decoherence of each
particle. We show that all initial states become separable after a
finite time, and we produce the explicit form of the separated
state. The result extends and amplifies an earlier result of
Di\'osi. We illustrate the general result by considering the case
in which the initial state is an EPR state (in which both the
positions and momenta of a particle pair are perfectly
correlated). This example clearly illustrates how the spreading
out in phase space produced by the environment leads to certain
disentanglement conditions becoming satisfied.

\end{abstract}



\maketitle

\def\A{{\cal A}}
\def\D{\Delta}
\def\H{{\cal H}}
\def\E{{\cal E}}
\def\p{\partial}
\def\la{\langle}
\def\ra{\rangle}
\def\ria{\rightarrow}
\def\x{{\bf x}}
\def\y{{\bf y}}
\def\k{{\bf k}}
\def\q{{\bf q}}
\def\p{{\bf p}}
\def\P{{\bf P}}
\def\r{{\bf r}}
\def\s{{\sigma}}
\def\a{\alpha}
\def\b{\beta}
\def\e{\epsilon}
\def\U{\Upsilon}
\def\G{\Gamma}
\def\om{{\omega}}
\def\Tr{{\rm Tr}}
\def\ih{{ {i \over \hbar} }}
\def\trho{{\rho}}

\def\au{{\underline \alpha}}
\def\bu{{\underline \beta}}
\def\pp{{\prime\prime}}
\def\id{{1 \!\! 1 }}
\def\half{{1 \over 2}}

\def\jjh{j.halliwell@ic.ac.uk}

\section{Introduction}

Much effort has recently been devoted to understanding the
properties of entangled quantum states. This effort is largely
driven by the emerging field of quantum computation and in
particular the desire to manipulate entangled states in a
practically useful way. However, another reason why the study of
entanglement is of interest concerns the question of emergent
classicality from quantum theory. Entanglement represents the
possibility of correlations which are greater than those
anticipated in classical theories. Hence, any account of emergent
classicality must explain how entanglement is lost. The
explanation of this is in fact reasonably simple and is closely
related to decoherence, the destruction of interference. The
purpose of this paper is to discuss the destruction of
entanglement in some simple systems and its connection to
decoherence.

A state of a bipartite system is said to be separable (or
disentangled) if it may be written in the form
\begin{equation}
\rho = \sum_i \ p_i \ \rho_i^A \otimes \rho_i^B
\label{1.1}
\end{equation}
where $p_i \ge 0 $ \cite{Wer}. Such a state describes essentially
classical correlations and can never violate Bell's inequalities.
However, it turns out to be surprisingly difficult to determine,
in general, whether a state may be written in this form. Peres
\cite{Per} made the very useful observation that a separable
$\rho$ remains a density operator under the operation of partial
transpose (transposition of one subsystem only). Hence, a
necessary condition for separability is that density operator
properties are preserved under partial transpose. This condition
was shown by Horodecki to be sufficient in the case of $ 2 \times
2 $ and $ 3 \times 3 $ dimensions \cite{Hor}, but generally not
otherwise. In the case of continuous variables the Peres-Horodecki
condition has a useful expression in terms of Wigner functions,
where (\ref{1.1}) becomes
\begin{equation}
W (p_1, x_1, p_2, x_2) = \sum_i \ p_i \ W_i^A (p_1, x_1) W_i^B (p_2, x_2)
\label{1.2}
\end{equation}
The Peres-Horodecki condition is then that $ W(p_1, x_1, p_2, x_2
) $ remains a Wigner function under $p_2 \rightarrow - p_2 $. It
has been shown by Simon that the condition is both necessary and
sufficient for the case when the Wigner function of a bipartite
system is Gaussian \cite{Sim}. Duan et al. \cite{Dua} considered a
different necessary and sufficient condition for the separability
of bipartite Gaussian states based on the variances of a class of
pairs of commuting EPR-like operators (of which $x_1 - x_2 $ and
$p_1 + p_2 $ are an example). (See Ref.\cite{Gie} for a discussion
of the connection between these condtions). There are undoubtedly
more conditions for Gaussian states.

A closely related idea is that of entanglement-breaking maps. This
is a map $\Phi$ acting on a subsystem $A$ such that
$(1_A\otimes\Phi)(\Gamma)$ is separable  for all choices of state
$\Gamma$ on $A\otimes B$, and for all finite-dimensional choices
of $B$. A theorem due to Horodecki and Shor then states that a map
$\Phi $ is entanglement-breaking if and only if it has the Holevo
form
\begin{equation}
\Phi (\rho ) = \sum_k \ R_k \ {\rm Tr} \left( F_k \rho \right)
\label{1.3}
\end{equation}
where $F_k$ are a set of POVMs and $ R_k$ are are set of density
operators which are independent of $\rho$ \cite{Sho,Rus,Hol}. Note
that this result refers to the dynamics of one of the subsystems
only. What is particularly interesting about this type of map, is
that appears naturally in the open system master equations of the
type frequently used in decoherence studies. In particular,
Di\'osi \cite{Dio} has recently considered the open system
dynamics described by the master equation
\begin{equation}
\dot \rho = \frac {i} {\hbar} [ H , \rho ] - \frac {D} {\hbar^2} [
x, [ x, \rho ] ] \label{1.4}
\end{equation}
This equation describes a particle coupled to a heat bath in the
limit of high temperature and negligible dissipation and is the
simplest equation used to describe decoherence. If we write the
solution to this equation as
\begin{equation}
\rho_t = \Phi ( \rho_0 )
\label{1.5}
\end{equation}
then Di\'osi has shown that this map becomes entanglement-breaking
after sufficient time has passed for the $P$-function to positive
\cite{Dio}. The time taken for this to happen is typically very
short and is essentially the same as the timescale required for
decoherence. Although what is particularly interesting is that
complete disentanglement occurs after a finite time, unlike
decoherence which, in the usual view of it, is asymptotic in time.
(See also the similar result by Di\'osi for spin systems in
Ref.\cite{Dio2}).

It is not hard to see that both decoherence and disentanglement
tend to be produced under the same circumstances. A simple
illustration of that fact is as follows. Consider first a
one-particle system in an initial superposition state
\begin{equation}
| \psi \rangle = | \psi \rangle + | \phi \rangle \label{1.6}
\end{equation}
and thus with density operator
\begin{equation}
\rho = | \psi \rangle \langle \psi| + | \psi \rangle \langle \phi
| + | \phi\rangle \langle \psi | + | \phi \rangle \langle \phi |
\label{1.7}
\end{equation}
Suppose now it is subject to evolution according to the master equation Eq.(\ref{1.4}),
with solution written in the form Eq.(\ref{1.5}). It is generally known that,
if the initial state (\ref{1.6}) is a superposition of localized position states,
evolution according to (\ref{1.4}) tends to kill the off-diagonal terms. That is,
we have
\begin{equation}
\Phi ( | \psi \rangle \langle \phi | ) \ \approx \ 0 \label{1.8}
\end{equation}
after a typically very short time. This means that the density operator becomes
essentially indistinguishable from the evolution of the mixed initial state,
\begin{equation}
\rho' = | \psi \rangle \langle \psi| + | \phi \rangle \langle \phi
| \label{1.9}
\end{equation}
This is the simplest account of decoherence
of a single particle coupled to an environment.

Now suppose we consider a two-particle system in the entangled state,
\begin{equation}
| \Psi \rangle = | \psi_1 \rangle \otimes |  \psi_2 \rangle + |
\phi_1 \rangle \otimes | \phi_2 \rangle
\end{equation}
with density operator
\begin{eqnarray}
\rho &=& | \psi_1 \rangle \langle \psi_1 | \otimes  | \psi_2
\rangle \langle \psi_2 | + | \psi_1 \rangle \langle \phi_1 |
\otimes  | \psi_2 \rangle \langle \phi_2 |
\nonumber \\
&+& | \phi_1 \rangle \langle \psi_1 | \otimes  | \phi_2 \rangle
\langle \psi_2 | + | \phi_1 \rangle \langle \phi_1 | \otimes  |
\phi_2 \rangle \langle \phi_2 |
\end{eqnarray}
If we now let both particles evolve according to the dynamics $
\Phi \otimes \Phi $, we find that once again the off-diagonal
terms go away, so for example,
\begin{equation}
\Phi ( | \psi_1 \rangle \langle \phi_1 | ) \ \approx \ 0
\end{equation}
and the density operator becomes indistinguishable from that obtained by the
initial state,
\begin{equation}
\rho = | \psi_1 \rangle \langle \psi_1 | \otimes  | \psi_2 \rangle
\langle \psi_2 | + | \phi_1 \rangle \langle \phi_1 | \otimes  |
\phi_2 \rangle \langle \phi_2 |
\end{equation}
which is separable. Hence, the mechanism that destroys
interference also destroys entanglement.

Another way to see why coupling to an environment will destroy
entanglement is to appeal to the fact that the property of
entanglement has an exclusive quality \cite{Woo}. Suppose $A$ is
entangled with $B$. Then if $B$ becomes entangled with a third
party $C$ it diminishes its entanglement with $A$. Hence an
environment coupling to one or both of the two particles in an
entangled state will cause one or both of them to become entangled
with the environment, thereby diminishing their entanglement with
each other.

The aim of this paper is to investigate the destruction of
entanglement through interacting with an environment, extending
and elaborating the earlier result of Di\'osi \cite{Dio,Dio2} and
others \cite{RaRe,Ven}

In Section 2 we consider the dynamics of a particle coupled to an
environment, concentrating on dynamics of the form Eq.(\ref{1.4}).
This is reasonably standard material but we write it in a form
which is most useful for studying disentanglement. In Section 3,
we consider the evolution of bipartite systems under the dynamics
of Section 2. We show that an arbitrary initial state achieves the
explicitly separated form (\ref{1.2}) after finite time. In
Section 4, we consider the evolution of the EPR state in the
presence of an environment. This simple example gives a clear
picture of how the various separability conditions come to be
satisfied as a result of interacting with the environment. We
summarize and conclude in Section 5.

\section{Evolution in the Presence of an Environment}

Before considering the evolution of entangled states in the
presence of a thermal environment, it is useful to consider first
the simplest case of a single particle coupled to a thermal
environment in the limit of high temperature and negligible
dissipation, with no external potential. The master equation
(\ref{1.4}) for the density matrix $ \rho (x,y) $ is,
\begin{equation}
 \frac {\partial \rho} { \partial t}
= \frac { i \hbar }{2 m} \left( \frac { \partial^2 \rho } {
\partial x^2} - \frac {\partial^2 \rho }{  \partial y^2} \right) -
\frac {D} {\hbar^2} (x-y)^2 \rho
\label{3.1}
\end{equation}
where $D = 2 m \gamma k T$. In the Wigner representation, the
corresponding Wigner function
\begin{equation}
W(p,x) = \frac {1}  {2 \pi \hbar} \int d \xi \ e^{-\ih p \xi}
\ \rho( x + \half \xi, x - \half \xi)
\label{3.1b}
\end{equation}
obeys the equation,
\begin{equation}
\frac {\partial W} {\partial t} = - \frac {p} {m} \frac {\partial W} {\partial x}
+ D \frac {\partial^2 W} {\partial p^2}
\label{3.2}
\end{equation}
Following Ref.\cite{AnHa}, this equation may be solved in the form
\begin{equation}
W_t (p,x) = \int dp_0 dx_0 \ K(p,x,t|p_0, x_0, 0 ) \ W_0 (p_0,x_0)
\label{3.3}
\end{equation}
where $K(p,x,t|p_0,x_0,0)$ is the Wigner function propagator, and
is given by
\begin{equation}
K = \exp \left[ - \alpha (p - p_{cl} ) - \beta ( x - x_{cl}  )
- \epsilon ( p - p_{cl} ) ( x - x_{cl}  ) \right]
\label{3.4}
\end{equation}
where $p_{cl}$ and $x_{cl}$ denote the classical evolution from
$p_0$, $x_0$ to time $t$,
\begin{equation}
p_{cl} = p_0, \quad x_{cl} = x_0 + \frac {p_0 t } {m}
\label{3.5}
\end{equation}
(For convenience, we ignore exponential prefactors unless
necessary).
 The coefficients $\alpha$, $\beta$ and $\epsilon$ are given by
\begin{equation}
\alpha = {1 \over Dt}, \quad \beta = { 3 m^2 \over D t^3}, \quad
\epsilon = - { 3 m \over D t^2}
\label{3.6}
\end{equation}
In fact, the general form of the propagator (\ref{3.4}) can be
used to describe the most general type of linear dynamics  --
arbitrary environment temperatures, non-negligible dissipation and
the inclusion of a harmonic oscillator potential -- for suitable
choices of $\alpha$, $\beta$, $\epsilon$ and $p_{cl}$, $x_{cl}$
\cite{AnHa}. More general dynamics are considered in another paper
\cite{Dod}.

With the simple change of variables $ x_0 \rightarrow x_0 - {p_0
t}/ {m} $ we may write,
\begin{eqnarray}
W_t (p,x) = \int & dp_0  dx_0  & \ \exp \left[ - \alpha (p - p_0 ) - \beta ( x - x_0  )
- \epsilon ( p - p_0 ) ( x - x_0  ) \right]
\nonumber \\
& \times & W_0' (p_0,x_0  )
\label{3.7}
\end{eqnarray}
where
\begin{equation}
W_0' (p_0, x_0) =
W_0 (p_0,x_0 -  {p_0 t} /{m} )
\label{3.8}
\end{equation}
This simple transformation is a linear canonical transformation,
which corresponds to a unitary transformation of the initial
state, so $W_0'$ is still a Wigner function. For decoherence and
disentanglement, the important aspects of the evolution are
contained in the convolution with the exponential function.
Following Di\'osi and Kiefer \cite{DiKi}, it is now very useful to
introduce the notation
\begin{equation}
\z = {p \choose x}
\label{3.9}
\end{equation}
and also to introduce a class of Gaussian phase space functions
\begin{equation}
g (\z; C) = \exp \left( - \half \z^T C^{-1} \z \right)
\label{3.10}
\end{equation}
The $ 2 \times 2 $ matrix $C$ is positive definite and $ | C | $ denotes its
determinant. The phase space function $ g (\z ; C) $ is a Wigner
function if and only if
\begin{equation}
| C | \ge \frac {\hbar^2} {4} \label{3.11}
\end{equation}
(This is essentially the uncertainty principle). A useful result
is the simple convolution property,
\begin{equation}
\int d^2 z \ g (\z_1 - \z ; C ) g (\z - \z_2 ; B ) = g ( \z_1 -
\z_2, C + B )
\label{3.12}
\end{equation}
We can use these Gaussians to compute smeared Wigner functions by
convolution:
\begin{equation}
\tilde W (\z ) = \int d^2 z' \ g ( \z - \z'; C ) W ( \z')
\label{3.13}
\end{equation}
Then it follows that the smeared Wigner function will be positive
if and only if (\ref{3.11}) holds. This is because the smeared Wigner
function is then equal to the overlap of two Wigner functions, for
which we have the result
\begin{equation}
\int dp dx \ W_{\rho_1} (p,x) W_{\rho_2} (p,x) = \frac {1} {2 \pi
\hbar} \ {\rm Tr} \left( \rho_1 \rho_2 \right)
\label{3.14}
\end{equation}
which is clearly always positive. For example, the $Q$-function,
which is always positive, is obtained in this way by smearing with
a minimum uncertainty Wigner function. In this notation the
propagation Eq.(\ref{3.7}) of the Wigner function is
\begin{equation}
W_t ( \z ) = \int d^2  z' \ g ( \z - \z '; A  ) \ W_0' ( \z ' )
\label{3.15}
\end{equation}
For the free particle without dissipation considered above we have
\begin{equation}
A = Dt \begin{pmatrix} 2 & t/m \\ t/m & 2t^2 / 3 m^2 \end{pmatrix}
\label{3.16}
\end{equation}
and therefore
\begin{equation}
| A | = \frac {D^2 t^4 } { 3 m^2}
\label{3.17}
\end{equation}
showing that the Wigner function tends to spread out with time.

Using the above description of the dynamics, it is straightforward
to compute the variances of $x$ and $p$ after a time $t$. They are
\begin{eqnarray}
(\Delta p)_t^2 &=& 2 D t + (\Delta p)_0^2
\label{3.18}\\
(\Delta x)_t^2 &=& \frac {2} {3} \frac {D t^3 } {m^2} + (\Delta p)_0^2 \frac{t^2} {m^2}
+ \frac {2} {m} \sigma (x,p ) + (\Delta x)_0^2
\label{3.19}
\end{eqnarray}
where
\begin{equation}
\s (x,p) = \half \langle \hat x \hat p +\hat p\hat x \rangle - \langle \hat x \rangle
\langle \hat p \rangle
\end{equation}
evaluated in the initial state.

\section{Evolution of Bipartite States in the Presence of an Environment}

Consider now the case of a two-particle system in an initially entangled
state. The two particles are not coupled to each other, but are each separately
coupled to a thermal environment, as described in the previous Section.

For our two-particle system, the Wigner evolution equation for the
Wigner function $ W(p_1, x_1, p_2, x_2)  = W( \z_1, \z_2 ) $ is,
\begin{equation}
\frac {\partial W} {\partial t} = - \frac {p_1} {m} \frac
{\partial W} {\partial x_1} - \frac {p_2} {m} \frac {\partial W}
{\partial x_2} + D \frac {\partial^2 W} {\partial p_1^2} + D \frac
{\partial^2 W} {\partial p_2^2}
\label{4.1}
\end{equation}
This equation may again be solved using propagators (with the unitary
part removed, as in Eqs. (\ref{3.7}), (\ref{3.8})).
The two-particle Wigner function at time $t$ is then given by
\begin{equation}
W_t ( \z_1, \z_2 ) = \int d^2 z_1' d^2 z_2 ' \ g ( \z_1 -\z_1'; A
) g (\z_2 - \z_2'; A ) \ W_0 '( \z_1', \z_2')
\label{4.2}
\end{equation}
where the matrix $A$ is given by Eq.(\ref{3.16}). We will show
that this evolves into the explicitly separable form (\ref{1.2})
after sufficient time has elapsed. The key idea is to write the
matrix $A$ in the propagator as
\begin{equation}
A = A_{1/4} + B
\label{4.3}
\end{equation}
where $A_{1/4}$ is the matrix of variances for a minimum uncertainty Wigner
function (whose explicit form will be given below). We
use the convolution property (\ref{3.12}) to write,
\begin{equation}
g ( \z_1 - \z_1'; A ) =\int d^2 \bar z_1 \ g (\z_1 - \bar \z_1 ; B
) g (\bar \z_1 - \z_1' ; A_{1/4} )
\label{4.4}
\end{equation}
and similarly with $ g( \z_2 - \z_2'; A)$. On the right-hand side,
the second Gaussian is a Wigner function because $ | A_{1/4} | =
\hbar^2/4 $. The first Gaussian with variance matrix $B$ will also
be a Wigner function if and only if
\begin{equation}
| B | = | A - A_{1/4} | \ge \frac {\hbar^2} {4} \label{4.5}
\end{equation}
We may now write the two-particle Wigner function at time $t$ as
\begin{equation}
W_t ( \z_1, \z_2 ) = \int d^2 \bar z_1 d^2 \bar z_2 \  \ g ( \z_1
-\bar \z_1; B ) g (\z_2 - \bar \z_2; B ) \ \tilde W_0 ( \bar \z_1,
\bar \z_2)
\label{4.6}
\end{equation}
where
\begin{equation}
\tilde W_0(\bar \z_1, \bar \z_2 ) = \int d^2 z_1' d^2 z_2' \ g (
\bar \z_1 - \z_1'; A_{1/4} ) g ( \bar \z_2 - \z_2'; A_{1/4} )\ W_0'
( \z_1', \z_2')
\label{4.7}
\end{equation}
and is positive, as explained above. This is the desired result. The Wigner function at time
$t$ is of the separable form (\ref{1.2}) as long as the matrix $B$ satisfies (\ref{4.5}).
This is because
the two Gaussians in (\ref{4.6}) are Wigner functions if Eq.(\ref{4.5}) holds, and the smeared Wigner
function $\tilde W_0 $ is positive and corresponds to the term $p_i$ in Eq.(\ref{1.2}).

We may also compare with the closely related result of Di\'osi
\cite{Dio}, who showed that a one-particle system achieves the
entanglement-breaking form (\ref{1.3}) after finite time, and
hence cannot be entangled with anything else. Tracing over
particle $2$ in Eq.(\ref{4.6}), we obtain
\begin{equation}
W_t (\z_1 ) = \int d^2 z_1 \ g( \z_1 - \bar \z_1 ; B ) \ \tilde W_0 (\bar \z_1 )
\label{4.8}
\end{equation}
where $\tilde W_0$ is the smeared Wigner function of the
one-particle system. This is clearly the Wigner transform of an
expression of the form (\ref{1.3}) because firstly, the Gaussian $
g(\z_1 - \bar \z_1; B ) $ is a class of Wigner functions
independent of the initial state so corresponds to $R_k$, and
secondly, $\tilde W_0$ is a $Q$-function so is of the desired form
$ {\rm Tr} \left( F_k \rho \right) $ with $F_k$ taken to be a
coherent state projector. Hence we completely agree with Di\'osi's
result. However, this result takes Di\'osi's result further in
that it gives the explicitly separated form of the disentangled
state.

Consider now the timescale on which the disentanglement
condition Eq.(\ref{4.5}) becomes satisfied,
where the matrix $A$ is given by Eq.(\ref{3.16}) and $A_{1/4}$ is still to be
chosen. Calculation of this timescale was reported (without explicit details)
in Refs.\cite{Dio,DiKi},
but we can extend this analysis in a small way. We also need to compare the
calculation with a result of the next section, so we give some of the details.

Since the calculation involves a small numerical calculation,
it is useful to define dimensionless variables $\bar t,\bar p,\bar x$, defined by
\begin{eqnarray}
t &=& \left( \frac {\hbar m} { D} \right)^{\half} \bar t
\label{4.9}
\\
p &=& (\hbar m D)^{\frac{1}{4}} \ \bar p
\label{4.10}
\\
x &=& \left( \frac {\hbar^3} { m D} \right)^{\frac{1}{4}} \bar x
\label{4.11}
\end{eqnarray}
In these dimensionless variables, we now take the matrix $A_{1/4}$ to be
\begin{equation}
A =  \begin{pmatrix} \sigma_0^2/\sqrt{2} & 1/2 \\ 1/2 & \sigma_0^{-2}/\sqrt{2} \end{pmatrix}
\label{4.12}
\end{equation}
Refs \cite{Dio,DiKi} made the particular choice $\sigma_0^2 = \sqrt{2}$ on the grounds
that this give a particularly robust evolution for certain sets of initial states
\cite{DiKi2}. For the moment, we leave it general. The condition Eq.(\ref{4.5}) now
becomes,
\begin{equation}
\frac {1} {3} \bar t^3 - \frac {2} {3} s \bar t^2 + \bar t - \frac {1} {s} \ge 0
\label{4.13}
\end{equation}
where $ s = \sigma_0^2 / \sqrt{2} $. The time at which this condition becomes satisfied
may be estimated by plotting the function in Maple for various values of $s$.
For the choice $\sigma_0^2 = \sqrt{2}$ we have $s = 1$,
and it is straightforward to then show that the
the condition is satisfied for values of $\bar t $ greater than about $1.39$.
This means,
\begin{equation}
t \ge  1.97 \times \left( \frac {\hbar m} { 2D} \right)^{\half}
\label{4.14}
\end{equation}
in agreement with Refs.\cite{Dio,DiKi}. (The factor of $2$ is included
because Refs.\cite{Dio,DiKi} used $D/2$ to denote what we denote by $D$).

However, if we tune the value of $\sigma_0$ to make the
timescale as short as possible, then numerical experiments show that the optimal
value is about $s=0.9$ which gives the slightly shorter timescale
\begin{equation}
t \ge  1.95 \times \left( \frac {\hbar m} { 2D} \right)^{\half}
\label{4.15}
\end{equation}
This improvement is clearly insignificant, but it does however
show that the Di\'osi-Kiefer choice is very nearly optimal.


\section{The EPR State}

Although the above results show disentanglement for general
initial states, it is of interest to look in detail at a particular
entangled state to see exactly how the entanglement goes away. We
therefore consider the entangled state first introduced by Einstein, Podolsky
and Rosen \cite{EPR}. Instead of the two canonical pairs $p_1, x_1$ and
$p_2, x_2$ it is very useful to use instead the rotated
coordinates,
\begin{eqnarray}
X &=& x_1 - x_2, \ \ \ \ K = \half ( p_1 - p_2)
\label{5.1}
\\
Q &=& ( x_1 + x_2), \ \ \ \ P = \half( p_1 + p_2 )
\label{5.2}
\end{eqnarray}
This is a canonical transformation, with the new canonical pairs
being $ K,X$ and $ P, Q $. However, we also have the important
relations,
\begin{equation}
[ X, P ] = 0, \ \ \ \ [ Q, K] = 0
\label{5.3}
\end{equation}
This simple observation is the basis of the EPR state, since it
means we may choose a representation in which $X$ and $P$ are
definite. In fact, EPR defined the state
\begin{equation}
\Psi (X, P ) = \delta (X) \delta (P)
\label{5.4}
\end{equation}
This state is not strictly normalizable. We therefore instead
consider Gaussian states which may be made arbitrarily close to
this state. In particular, we consider a Gaussian state, which, in
the rotated coordinates has Wigner function
\begin{equation}
W(K,X,P,Q) = \exp \left(- \frac {K^2}{ 2 \s_K^2}
- \frac { X^2} {2 \s^2_X} - \frac { P^2}
{ 2 \s^2_P}   -  \frac { Q^2} { 2 \s_Q^2}
\right)
\label{5.5}
\end{equation}
The widths $\s_X$ and $\s_P$ may be chosen to be arbitrarily
small, but for this to be a Wigner function, the remaining widths
must satisfy
\begin{equation}
\s^2_K \s^2_X \ \ge \ \frac {\hbar^2} {4}, \ \ \ \ \s^2_P \s^2_Q \
\ge \ \frac {\hbar^2} {4}
\label{5.6}
\end{equation}
(Because the transformation from the original coordinates to
rotated ones is a linear canonical transformation, it preserves
Wigner function properties, so the conditions to be a Wigner
function have the same form as in the original coordinates.)

As an aside, we remark that Bell has noted that Gaussian states
have a positive Wigner function,
and therefore have hidden variable interpretation, so cannot
violate Bell's inequalities \cite{Bell}. This perhaps suggests
that they do not have any significant
entanglement properties. This is,
however, in terms of observables local in phase space quantities.
It has been pointed out that there are other observables, in particular
the parity operator, in terms
of which Gaussian states do violate Bell's inequalities \cite{BaWo}.
Hence it is of interest to study entanglement of Gaussians.

In terms of the variables (\ref{5.1}), (\ref{5.2}), the
Peres-Horodecki \cite{Per,Hor,Sim} condition for disentanglement is the condition
that $W(K,X,P,Q)$ must remain a Wigner function when $P$ and $K$
are interchanged, that is, that
\begin{equation}
\tilde W (K,X,P,Q) = \exp \left(  - \frac {P^2}{ 2 \s_K^2}- \frac { X^2} {2 \s^2_X} - \frac
{ K^2} { 2 \s^2_P}  -  \frac { Q^2} { 2
\s_Q^2}  \right)
\label{5.7}
\end{equation}
is a Wigner function. The conditions for this to be a Wigner
function are,
\begin{equation}
\s^2_X \s^2_P \ \ge \ \frac {\hbar^2} {4}, \ \ \ \ \s^2_Q \s^2_K \
\ge \ \frac {\hbar^2} {4}
\label{5.8}
\end{equation}
which is different to (\ref{5.6}). In particular, the first of
these relations will not be satisfied when $\s_X $ and $\s_P$ are
chosen to be very small, which is the EPR case. This is generally
the case for entangled states -- they fail to remain Wigner functions
under interchange of $P$ and $K$ because they are then too strongly
peaked in phase space an violate the uncertainty principle.

To watch the destruction of entanglement, we will work with the
disentanglement condition put forward by Duan et al. \cite{Dua},
rather than the Peres-Horodecki condition. They wrote down
a class of necessary and sufficient conditions for a Gaussian
state to be disentangled. In terms of
dimensionless variables $\bar K, \bar X, \bar P, \bar Q$
(defined as in Eqs. (\ref{4.9})--(\ref{4.11})), one of those conditions is
\begin{equation}
(\Delta \bar X)^2 + 4 ( \Delta \bar P)^2 \ge 2
\label{5.9}
\end{equation}
This is clearly not satisfied for the EPR initial state
Eq.(\ref{5.5}), but it is of interest to see how it becomes
satisfied under evolution in the presence of an environment. The
key point here is that the condition Eq.(\ref{5.9}) means that the
state is reasonably spread out in phase space, and this phase
space spreading is precisely what evolution in the presence of an
environment produces, so we expect that the condition will become
satisfied after a short period of time.

In the presence of an environment, the initial state will evolve
according to Eq.(\ref{4.1}).
However, since the regularized EPR state (\ref{5.5}) factors in
terms of the rotated coordinates (\ref{5.1}), (\ref{5.2}), it is more useful to use those,
and in terms of them, the Wigner equation is
\begin{equation}
\frac {\partial W} {\partial t} = - \frac {2 P} {m} \frac {\partial
W} {\partial Q} - \frac {2 K} {m} \frac {\partial W} {\partial X} +
D \frac {\partial^2 W} {\partial P^2} + D \frac {\partial^2 W}
{\partial K^2}
\label{5.10}
\end{equation}
Interestingly, the dynamics also factors in the rotated variables
and is essentially the same as the single-particle dynamics.
Using Eqs.(\ref{3.18}) and (\ref{3.19}), it is then easily seen that
\begin{eqnarray}
(\Delta P)_t^2 &=& 2 D t + \sigma_P^2
\label{5.11}\\
(\Delta X)_t^2 &=& \frac {8} {3} \frac {D t^3 } {m^2} + \frac{4} {m^2} \sigma_K^2 t^2
 + \sigma_X^2
\label{5.12}
\end{eqnarray}
Inserting these expressions in the condition Eq.(\ref{5.9}), we find that,
in dimensionless variables, it is satisfied as long as
\begin{equation}
\frac{8} {3} \bar t^3 + 4 c \bar t^2 + 8 \bar t + \frac {1} {4c} \ge 2
\label{5.13}
\end{equation}
Here, $c$ is the dimensionless form of $\sigma_K^2$ and we have used the
uncertainty principle to eliminate $\sigma_X^2$. The interesting case is
that in which $\sigma_X^2$ is very small, but we cannot set it to zero
because then $ \sigma_K^2$ would be infinite. We have also set
$\sigma_P^2 = 0$, which represents the most extreme case. (We can do
this because $\sigma_Q^2$, its conjugate width, does not appear).

The polynomial in Eq.(\ref{5.13}) may be plotted using Maple for various values of $c$.
We have found from these plots that for all values of $c$ the condition becomes
satisfied for values of $\bar t$ greater than about 0.19, that is, for
\begin{equation}
t \ge 0.27 \times \left( \frac {\hbar m} { 2D} \right)^{\half}
\label{5.14}
\end{equation}
This is the time after which the EPR state becomes disentangled, according
the condition Eq.({\ref{5.9}). This time is actually a lot shorter than
the timescale Eq.(\ref{4.15}), hence is safely consistent with the general
result. The timescale Eq.(\ref{4.15}) is the time for {\it any}
initial state to become disentangled whereas the timescale Eq.(\ref{5.14})
is only for the EPR state. This actually indicates that the EPR state is
perhaps not a very entangled state, at least in terms of phase space
variables.



\section{Discussion}

We have shown, in a variety of different ways, how open system dynamics destroys
entanglement in a system consisting of two entangled particles. Entanglement
is destroyed by the same mechanism that destroys interference.
In particular, we have shown that, under a simple open system dynamics,
any initial two-particle state achieves the explicitly disentangled form Eq.(\ref{1.2})
after a short, finite time. We illustrated this general result with the particular
case of the EPR state.

The dynamics employed here are those of the free particle coupled to
a bath of oscillators, in the limit of negligible dissipation. It is of interest
to extend the analysis to include dissipation, a potential, and also to include
interactions between the particles which will then compete with the disentangling
effects of the environment. This is considered in another paper \cite{Dod}.

As previously noted \cite{Dio,Dio2}, it is striking that complete disentanglement is achieved
in finite time, whereas the interference terms in the density matrix tend
to take infinite time to completely go away. That is, one is inclined to say
that complete decoherence takes an infinite time.
Although note that the word ``decoherence'' is used in a variety of different ways.

This perhaps suggests that the exact vanishing of the off-diagonal
terms in the density matrix is perhaps too strong a condition for determining
when a quantum system is essentially classical. It is, for example, often
suggested that positivity of the Wigner function is a useful condition
characterizing quasiclassicality, because this then means that the one-particle
system is a hidden variables theory. Significantly, the Wigner function typically
becomes positive after a finite time in open system dynamics. Hence, a reasonable
meaning to attach to the word decoherence is that it is the situation under
which prediction for all variables may be described by a hidden variables theory.
These ideas will be explored in future publications.
See Ref.\cite{Hal1} for related discussions.

\section{Acknowledgements}

We are very grateful to Lajos Di\'osi for useful conversations.
P.D. was supported by PPARC.

\bibliography{apssamp}

\end{document}